\documentclass[12pt]{iopart}

\usepackage{graphicx}
\usepackage{color}
\usepackage{verbatim}
\newcommand*{\wn}{cm$^{-1}$}
\newcommand*{\Hmol}{H$_{2}$}
\newcommand*{\Dmol}{D$_{2}$}
\newcommand*{\Tmol}{T$_{2}$}
\newcommand*{\wprobe}{$\omega_\mathrm{probe}$}
\newcommand*{\wpump}{$\omega_\mathrm{pump}$}
\newcommand*{\wsignal}{$\omega_\mathrm{anti-Stokes}$}

\begin{document}

\paper{CARS spectroscopy of the $(v=0\rightarrow 1)$ band in $\rm{T_2}$}

\author{M. Schl\"{o}sser$^1$, X. Zhao$^1$, M. Trivikram$^2$, W. Ubachs$^2$,\\ E. J. Salumbides$^2$}
\address{$^1$ Tritium Laboratory Karlsruhe, Institute of Technical Physics,\\ Karlsruhe Institute of Technology, Hermann-von-Helmholtz-Platz 1, 76344 Eggenstein-Leopoldshafen, Germany}

\address{$^2$ LaserLaB and Department of Physics and Astronomy, Vrije Universiteit,
\\De Boelelaan 1081, 1081 HV Amsterdam, The Netherlands}
\ead{magnus.schloesser@kit.edu; e.j.salumbides@vu.nl}

\def\theabstract{
Molecular hydrogen is a benchmark system for bound state quantum calculation and tests of quantum electrodynamical effects.
While spectroscopic measurements on the stable species have progressively improved over the years, high-resolution studies on the radioactive isotopologues $\rm{T_2}$, $\rm{HT}$ and $\rm{DT}$ have been limited.
Here we present an accurate determination of \Tmol\ $Q(J=0-5)$ transition energies in the fundamental vibrational band of the ground electronic state, by means of high-resolution Coherent Anti-Stokes Raman Spectroscopy.
With the present experimental uncertainty of $0.02\,\rm{cm^{-1}}$, which is a fivefold improvement over previous measurements, agreement with the latest theoretical calculations is demonstrated.
}


\submitto{\JPB}



\begin{abstract}
\theabstract
\end{abstract}
\maketitle

\section{Introduction}

Molecular hydrogen and its isotopologues are benchmark systems where the most refined quantum chemical theories can be applied.
Its energy level structure can be calculated to the highest accuracy of any neutral molecule, which include the Born-Oppenheimer (BO) contribution, nonrelativistic BO corrections as well as relativistic and radiative (quantum electrodynamic or QED) contributions~\cite{Piszczatowski2009,Komasa2011}.
Adiabatic and nonadiabatic effects are prominent in molecular hydrogen due to the lightness of the nuclei, and thus analogous studies on the different isotopologues provide a handle for a better understanding of these contributions to the level energies.

The ground state level energies and transitions in \Hmol, HD and \Dmol\ have been determined at ever increasing accuracies, with the recent accurate studies employing cavity-enhanced techniques ( e.g.~\cite{Maddaloni2010,Campargue2012,Kassi2011,Kassi2012,Tan2014} ) or pulsed molecular beam spectroscopies~\cite{Salumbides2011,Dickenson2013}.
The dissociation energy $D_0$, a benchmark quantity in quantum molecular calculations, have also been determined to a high accuracy~\cite{Liu2009, Liu2010, Sprecher2010} by combining the results of several precision spectroscopies.  
The comparison of the results from measurements and calculations enable precision tests of QED, and even test of new physics beyond the Standard Model~\cite{Ubachs2016}.

While numerous precision spectroscopic studies on the stable molecular hydrogen isotopologues have been performed, the tritium-bearing species HT, DT and \Tmol\ are not as well studied because of the limited access to radioactive tritium.
The comprehensive classical spectroscopy of Dieke~\cite{Dieke1958} on all isotopologues of molecular hydrogen included tritium-containing species.
Laser Raman spectroscopic studies were carried out on the pure rotational (0-0) and fundamental vibrational (1-0) bands by Edwards, Long and Mansour~\cite{Edwards1978} for \Tmol\ and later for HT and DT~\cite{Edwards1979}.
Veirs and Rosenblatt~\cite{Veirs1987} also performed Raman spectroscopy on all molecular hydrogen species, however, they reported discrepancies with Ref.~\cite{Edwards1978,Edwards1979} that exceeded the claimed uncertainties of those previous determinations.
Chuang and Zare~\cite{Chuang1987} performed high-resolution measurements on the (1-0), (4-0) and (5-0) bands of HT using photoacoustic techniques.
They pointed out discrepancies when comparing to the \emph{ab initio} calculations by Schwartz and LeRoy~\cite{Schwartz1987} that may stem from the treatment of nonadiabatic corrections.

The molecular structure of the tritiated hydrogen isotopologues in the electronic ground state is also relevant in experiments which aim at the determination of the mass of the neutrino, like the Karlsruhe Tritium Neutrino experiment (KATRIN) \cite{Design-Report}. The mass is deduced from high-resolution electron spectroscopy of the decay electrons from the molecular beta-decay of tritium, $\rm{T_2} \to \rm{^{3}HeT}^+ + e^- + \bar{\nu}_e$. After the decay, the helium-tritium ion remains in an excited electronic, vibrational and/or rotational state \cite{Bodine2015}. This internal excitation energy reduces the kinetic energy carried away by the electron, which acts as the central kinematic observable in the neutrino mass measurements. The probability of excitation of these states is currently only accessible via theory \cite{Saenz00, DossFinalState}. These calculations crucially depend on the energy levels of the initial states of $\rm{T_2}$ and final states of $^3\rm{HeT}^+$.

In this contribution, we present our determination of the $Q$-branch transitions of the \Tmol\ fundamental band $(v=0\rightarrow 1)$ in the ground electronic state.
We employed Coherent Anti-Stokes Raman spectroscopy (CARS) \cite{Clark1988} using narrowband pulsed lasers on a low-pressure gas cell containing tritium.
Comparison with previous experimental determinations demonstrate agreement with Ref.~\cite{Veirs1987}, while comparison with calculations in Ref.~\cite{Pachucki2015} indicate that relativistic and QED contributions are smaller than the present measurement accuracy.

\begin{figure}
	\centering
		\includegraphics{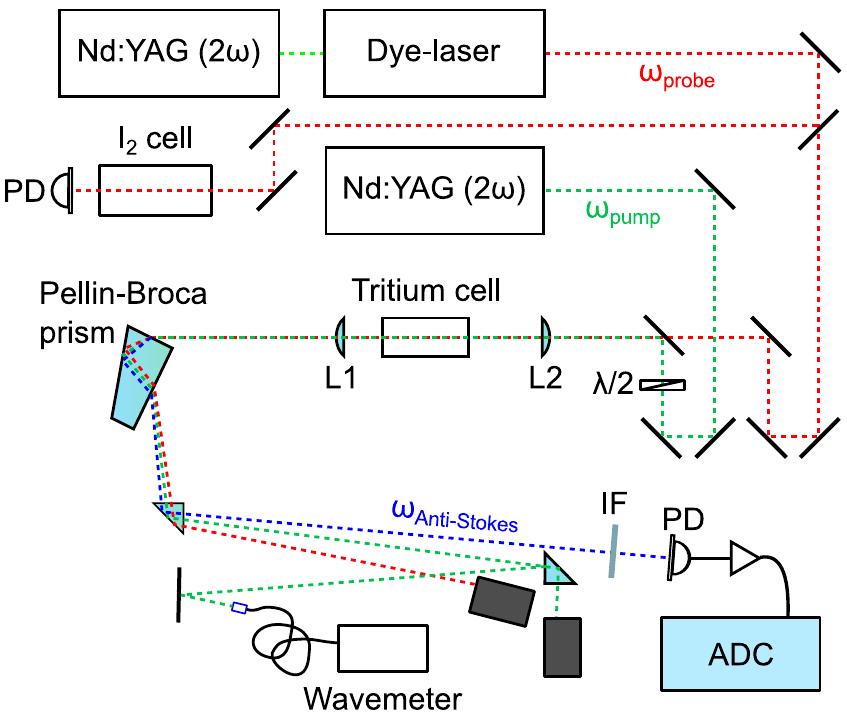}
	\caption{Schematic diagram of CARS setup. $\lambda/2$: half-wave plate; L1, L2: lenses; IF: interference filter; PD: photodiode.
}
	\label{fig:Setup}
\end{figure}

\section{Experiment}
For the high resolution CARS measurements a new tritium compatible cell was developed based on the concept of the standard Raman cells at the Tritium Laboratory Karlsruhe (TLK) \cite{Taylor2001, Sturm2010}. Tritium-compatibility is guaranteed by employing well-proven materials only, such as stainless steel or certain fused silica windows with metal-to-window sealings. It should be noted that rubber- or plastic-based materials cannot be used due to deterioration by radiochemical reactions induced by the beta-emitter. The inner volume of the cell should be a small as possible to reduce the activity of the tritium gas sample at a certain pressure. On the other hand the length of the cell needs to be long enough, so that the energy density of the laser pulses at the window surface does not exceed the damage threshold. 
In our design a good compromise was found at a length of $80\,\rm{mm}$, which resulted in a total inner cell volume of $4\,\rm{cm^3}$. Tritium gas was prepared at high purity by performing displacement gas chromatography and gettering on uranium beds at the TLK \cite{Doerr2008} and filled into the cell at about $2.5\,\rm{mbar}$. This corresponds to an activity of less than $1\,\rm{GBq}$ which is the legal limit for handling tritium outside of a licensed lab. The composition of the sample was measured by Raman spectroscopy \cite{Schloesser_TLK-Raman_2015}: $\rm{T_2}:93.4\%$, $\rm{DT}:4.9\%$, $\rm{HT}:1.4\%$, $\rm{H_2}:0.2\%$ and $\rm{D_2}:0.1\%$. Subsequently, the cell was decontaminated to allow transporting to the CARS setup at LaserLab Amsterdam.

\begin{figure}
	\centering
		\includegraphics{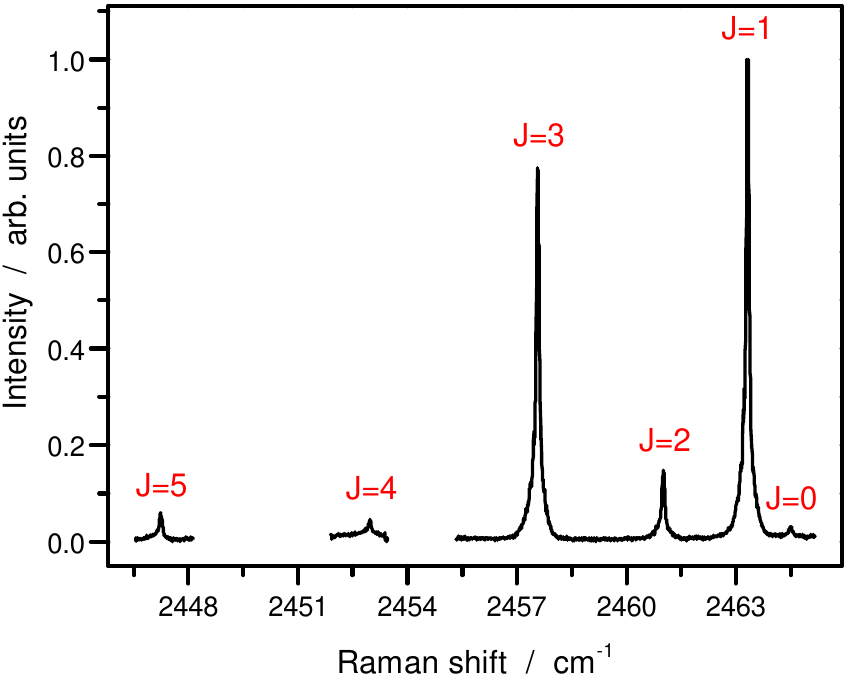}
	\caption{High-resolution spectrum of $\rm{Q}$-branch in the fundamental band of $\rm{T_2}$. The lines shown were recorded at pump/probe pulse energies of 5 mJ, except for $J=4$ which was recorded at 7.5 mJ.}
	\label{fig:Spectrum}
\end{figure}

A schematic diagram of the setup is shown in Fig.~\ref{fig:Setup}. The frequency-doubled output, at 532 nm, of an injection-seeded Nd:YAG pulsed laser (Spectra Physics 250-10 PRO, 10-Hz repetition rate) was used as the pump radiation \wpump. 
The tunable probe beam \wprobe, at $\sim 612$ nm, is the output of a pulsed dye laser (Lioptec LiopStar) running on a mixture of Rhodamine B and Rhodamine 640 that is pumped by another Nd:YAG pulsed laser. The pump and probe beams are spatially combined and subsequently focused by a lens L2, focal length $f=30$ cm, into the tritium cell and collimated by another lens L1, focal length $f=10$ cm, after the cell. The probe, pump and anti-Stokes signal $\omega_\mathrm{anti-Stokes} = 2\omega_\mathrm{pump} - \omega_\mathrm{probe}$ beams are then dispersed by a Pellin-Broca prism, where the signal beam ($\sim 470$ nm) passes through an optical interference filter (bandwidth 10 nm) before collection onto a photodetector. The temporal overlap of the pump and probe pulses is optimized by changing the trigger timings of the Nd:YAG lasers.

During the measurement, the wavelength of \wpump\ is measured continuously (around 18789.0302(10) \wn) using a wavemeter (High Finesse Angstrom WSU-30).
The probe radiation \wprobe\ wavenumber is scanned, while simultaneously a portion of \wprobe\ is diverted into an I$_2$ absorption cell, where the Doppler-broadened absorption lines of the latter are used as calibration reference \cite{Gerstenkorn1977}. The iodine calibration lines are shown together with a selected Raman line in Fig. \ref{fig:PulseEnergy}.
The wavenumber calibration of \wpump\ is estimated to be better than $0.002$ \wn, while the calibration accuracy of \wprobe\ is estimated at $0.01$ \wn.

\section{Results and Discussion}

The observed $Q(0) - Q(5)$ transitions $(v=0\rightarrow 1, \Delta J=0)$, plotted against the Raman shift (\wsignal), are shown in Fig.~\ref{fig:Spectrum}.
Good signal-to-noise ratio is obtained even for the weakest lines for 7.5 mJ pulse energy for both the pump and probe beams, where the linewidths are observed to be $\sim0.15$ \wn.
The relative intensities of the CARS signal follow the expected behaviour of $I_{\rm{CARS}} \propto N^2$ with $N$ being the number density by considering Placzek-Teller Raman line strength coefficients and a room temperature partition function \cite{Slenczka1988}.

\begin{figure}
	\centering		
	  \includegraphics{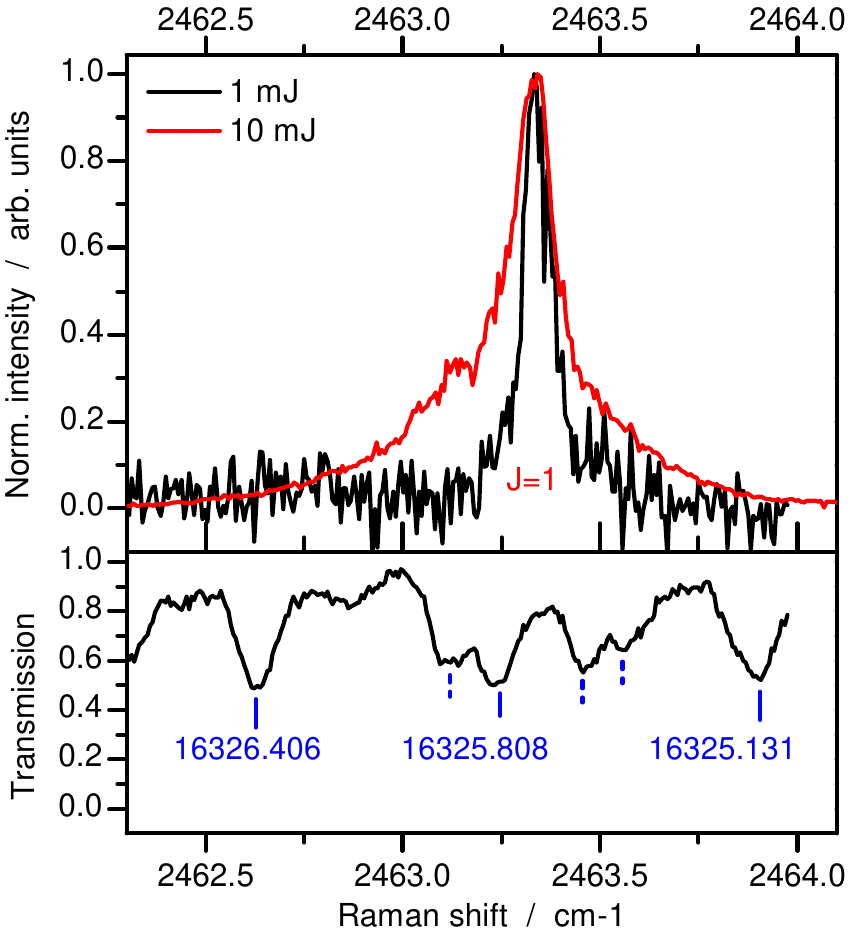}
	\caption{ Top: $Q(1)$ transition recorded at $1$ and $10\,\rm{mJ}$ for the pump and probe beams.
Bottom: Transmission through $\rm{I_2}$ calibration cell. Blue lines indicate I$_2$ absorption lines which have been included in the calibration of the tuneable probe laser (in $\rm{cm^{-1}}$). Note, that the wavenumber axis of the iodine spectrum is given by the wavenumber of the pump ($18789.0302(10)\,\rm{cm^{-1}}$) - Raman shift. }
	\label{fig:PulseEnergy}
\end{figure}

The Doppler width of the T$_2$ $Q$ transitions at room temperature is $0.014$ \wn.
In order to estimate the effect of ac-Stark broadening, the $Q(1)$ transition was recorded at lower pulse energies of 1 mJ for both pump and probe beams, see Fig.~\ref{fig:PulseEnergy}.
Even at these reduced pulse energies, the linewidth is observed to be $\sim0.1$ \wn, which we attribute to be mainly limited by the probe laser ($\omega_\mathrm{probe}$) instrument bandwidth.
This is confirmed by the fitted linewidths of the I$_2$ calibration resonances of $\sim 0.08$ \wn, used for calibrating \wprobe, while the expected I$_2$ Doppler width is $\sim 0.03$ \wn.
The instrument linewidth of the pump laser \wpump\ is estimated to be $\sim 0.005$ \wn, and thus has a minor contribution to the \Tmol\ linewidths.

The ac-Stark effect shifts the $v=0$ and $v=1$ energy levels due to effects of the polarizability derivative and by a smaller contribution of the polarizability anisotropy \cite{Rahn1980, Farrow1982}. Both these properties are different in ground and excited states and the resulting ac-Stark coefficient for H$_2$ are found: $\Delta\alpha = 20\,\rm{MHz/GHz\cdot cm^2}$ for the polarizability derivative and $\Delta\gamma = 9\,\rm{MHz/GHz\cdot cm^2}$ for the polarizability anisotropy. Since this property is different in the ground- and excited state, a shift of the resulting Raman lines appears. Dyer and Bischel \cite{Dyer1991} showed that the experimentally determined ac-Stark shifts are in perfect agreement with \emph{ab initio} calculated polarizabilities and that the ac-Stark effect for hydrogen always leads to a red-shift of the Raman line. Using the formulas of Dyer and Bischel with polarizability values for $\rm{T_2}$ by Schwartz and Le Roy \cite{Schwartz1987}, one obtains a shift which is $40\%$ smaller than that of $\rm{H_2}$. 
When calculating the expected shift for a CARS signal with typical pulses (pump/probe pulse energy $E=10\,\rm{mJ}$, pulse duration $\tau=8\,\rm{ns}$, beam diameter in focus $d_{\rm{focus}}=35\,\rm{\mu m}$) we should expect a shift of about $0.1\,\rm{cm^{-1}}$. A shift like this should have been visible in Fig. \ref{fig:PulseEnergy}. The effect on our spectrum is less a shift, but more a broadening towards the red (lower wavenumber) side of the CARS resonance. This observation may be ascribed to a difference in our CARS experiment as compared to the dedicated setup by Dyer and Bischel with a well-controlled auxiliary laser for quantifying the ac-Stark effect. In our approach temporal and spatial integration over the power densities has to be considered for the quantitative assessment of the ac-Stark phenomenon, leading to the observed red-side broadening as demonstrated by Li, Yang and Johnson \cite{Li1985}. As we measured the line profiles for energies for 1 to 10 mJ while accounting for the asymmetric broadening, we did not observe a shift above the uncertainty of the probe laser calibration of $0.01\,\rm{cm^{-1}}$.  

May et al. \cite{May1961} developed a model for the collisional shift in $\rm{H_2}$ as a function of pressure. They measured the linear shift coefficient of the Q-branch to be about $\sim0.003$ \wn/amagat at room temperature (See also Rahn et al. \cite{Rahn1990}). They predicted that the shift is proportional to the fundamental vibrational transition $\omega_e$ and thus to the inverse square root of the reduced mass $(\propto\mu^{-1/2})$ - a behavior which has been verified by Looi et al. \cite{Looi1978} when comparing shifts in \Dmol\ and \Hmol. Therefore, the linear shift coefficient in \Tmol\ is expected be about 1.7 smaller as compared to that of \Hmol. At a pressure of 2.5 mbar, we estimate the collision shift to be negligible ($<1\times10^{-5}$ \wn). 

The accuracy of the present line position determination is then limited by the line fitting uncertainty and by the Doppler-broadened I$_2$ calibration reference. We provide a conservative estimate of $0.02$ \wn\ for the transitions measured.
Multiple measurements of the $Q(1)$ transition demonstrate reproducibility of the line positions to within $0.01$ \wn\ at the highest pulse energy of 10 mJ, at lower pulse energies the reproducibility increases.

The fitted line positions of the $Q$ transitions are listed in Table~\ref{table:transitions}. The transition values from the previous determination of Veirs and Rosenblatt~\cite{Veirs1987}, as well as the calculations of Schwartz and LeRoy~\cite{Schwartz1987} and by Pachucki and Komasa~\cite{Pachucki2015} are also included in the table.
There is good agreement with the previous experimental values of Ref.~\cite{Veirs1987}, with the present results representing a fivefold improvement.
The calculation of Schwartz and LeRoy~\cite{Schwartz1987} include early estimates of the relativistic and radiative corrections to the potentials, where they estimate a $0.01$ \wn\ contribution to the level energies.
They also evaluate the nonadiabatic corrections by an isotopic scaling procedure and estimate a total uncertainty of $0.015$ \wn\ for the level energies of low rotational $J$ states.
The calculations of Pachucki and Komasa~\cite{Pachucki2015} only dealt with nonrelativistic contributions including nonadiabatic corrections for which they assign an uncertainty of $10^{-7}$ \wn, which is effectively exact for our purpose.
Although Ref.~\cite{Pachucki2015} does not include relativistic and radiative (QED) contributions, good agreement is found with the present results.
This suggests that relativistic and QED corrections are below the present experimental uncertainty of $0.02$ \wn.
Since the corresponding relativistic and QED corrections for $Q$ transitions is $\sim0.002$ \wn\ in D$_2$ \cite{Komasa2011}, the present agreement with Ref. ~\cite{Pachucki2015} is reasonable as it is expected that the analogous corrections are less for T$_2$.    
It is worth noting that the calculation by Schwartz and Le Roy~\cite{Schwartz1987} agrees to within $0.004$ \wn\ when compared to Ref.~\cite{Pachucki2015}.

\begin{table}
\begin{center}
\caption{
Transition energies of the $Q-$branch in the fundamental band ($v=0\rightarrow 1$) of T$_2$ extracted from measurement with lowest possible pump/probe pulse energies at sufficient high signal-to-noise ratio ($J=0,1,2,3,5$: $5\,\rm{mJ}$; $J=4$: $7.5\,\rm{mJ}$).
The present results are compared to the previous experimental determination in Ref.~\cite{Veirs1987}
as well as calculations in Ref.~\cite{Schwartz1987,Pachucki2015}.
See text for the discussion of uncertainties in the \emph{ab initio} calculations.
The energies are given in \wn.}\label{table:transitions}
\begin{small}
\begin{tabular}{ccccc}
\noalign{\smallskip}\hline\noalign{\smallskip}
$J$ & this work & Ref.~\cite{Veirs1987} & Ref.~\cite{Schwartz1987} & Ref.~\cite{Pachucki2015}\\
\noalign{\smallskip}\hline\noalign{\smallskip}
0	& 2464.50\,(2)	& 		& 2464.498\,(10)	& 2464.502\\
1	& 2463.33\,(2)	& 		& 2463.343\,(10)	& 2463.346\\
2	& 2461.01\,(2)	& 2461.0\,(1)	& 2461.034\,(10)	& 2461.037\\
3	& 2457.57\,(2)	& 2457.5\,(1)	& 2457.575\,(10)	& 2457.579\\
4	& 2452.97\,(2)	& 2452.0\,(1)	& 2452.976\,(10)	& 2452.980\\
5	& 2447.23\,(2)	& 2447.3\,(1)	& 2447.245\,(10)	& 2447.249\\
\noalign{\smallskip}\hline\noalign{\smallskip}
\end{tabular}
\end{small}
\end{center}
\end{table}

The previous measurements of Edwards, Long and Mansour~\cite{Edwards1978} deviate by $-0.2$ \wn\ in comparison to the present results on average, despite their claimed accuracy of $0.005$ \wn.
As the same disagreement persists when comparing Ref.~\cite{Edwards1978} with the experimental results of Veirs and Rosenblatt~\cite{Veirs1987}, we believe that determinations of Edwards, Long and Mansour~\cite{Edwards1978} may have unrecognized systematic errors.

\section{Conclusions}

We have determined the transition energies of the $Q(0)-Q(5)$ lines in the fundamental band ($v=0\rightarrow 1$) of the ground electronic state of molecular tritium by employing CARS spectroscopy.
The present results are in good agreement with Veirs and Rosenblatt~\cite{Veirs1987} and represent a five-fold improvement over their previous experimental determination.
Agreement with the most accurate nonrelativistic calculation~\cite{Pachucki2015} implies that relativistic and radiative effects for these transitions are below the present accuracy of $0.02$ \wn, but may manifest in planned measurements using a narrowband probe laser source.
Our present results on T$_2$, with the addition of spectroscopies on tritiated species HT and DT, expand our ongoing program on testing the most accurate quantum chemical theory.

\ack
We thank Tobias Falke, David Hillesheimer, Stefan Welte and J\"urgen Wendel of the Tritium Laboratory Karlsruhe for the welding of the cell, preparation of the tritium gas, the filling of the cell and the legal support for the cell transport.
WU thanks the European Research Council for an ERC-Advanced grant (No 670168).
The research leading to these results has received funding from LASERLAB-EUROPE (grant agreement no. 654148, European Union’s Horizon 2020 research and innovation programme).
\section*{References}
	\bibliography{t2-cars-biblio}
  \bibliographystyle{iopart-num}

\end{document}